\documentclass[onecolumn]{aastex631}
\usepackage{amsmath}
\usepackage{csvsimple}

\begin{document}

\title{Elastic Planetoids}

\author[0009-0002-6286-4499]{Bartosz Żbik}
\affiliation{Institute of Theoretical Physics, Jagiellonian University in Krak\'{o}w, Poland}

\author[0000-0002-0532-4907]{Andrzej Odrzywołek}
\affiliation{Institute of Theoretical Physics, Jagiellonian University in Krak\'{o}w, Poland}

\correspondingauthor{Andrzej Odrzywołek}
\email{andrzej.odrzywolek@uj.edu.pl}

\begin{abstract}
Modeling the internal structure of self-gravitating solid and liquid bodies presents a challenge, as existing approaches are often limited to either overly simplistic constant-density approximations or more complex numerical equations of state.  We present a detailed analysis of a tractable and physically motivated model for perfectly elastic, spherically symmetric self-gravitating bodies in hydrostatic equilibrium. The model employs a logarithmic equation of state (logotropic EOS) with a non-zero initial density and constant bulk modulus. Importantly, scaling properties of the model allow all solutions to be derived from a single, universal solution of an ordinary differential equation, resembling the Lane-Emden and Chandrasekhar models. The model provides new insights into stability issues and reveals oscillatory asymptotic behavior in the mass-radius relation, including the existence of both a maximum mass and a maximum radius. We derive useful, simple analytical approximations for key properties, such as central overdensity, moment of inertia, binding energy, and gravitational potential, applicable to small, metallic bodies like asteroids and moons. These new approximations could aid future research, including space mining and the scientific characterization of small Solar System bodies.
\end{abstract}



\section{Introduction} \label{sec:intro}

The hydrostatic equilibrium of spherically symmetric, self-gravitating bodies is a cornerstone of research and education in astronomy, astrophysics, and planetary science. Simple analytical models are particularly valuable because they are widely applicable, easy to understand, and facilitate quick calculations and discussions grounded in established concepts. A prime example is the Lane-Emden equation \citep{Lane1870On}, which, with its power-law Equation Of State (thereafter EOS), describes a wide range of objects, ranging from constant density planetoid models and gas giants \citep{10.1093/mnras/stx164} to stars via e.g. the Eddington model \citep{Eddington} and even globular clusters \citep{Plummer}. Its general relativistic generalization \citep{1964ApJ...140..434T} has been applied to model cosmological voids \citep{2009PhRvD..80j3515O}. The Chandrasekhar white dwarf model \citep{1935MNRAS..95..207C, Kippenhahn2012}, based on the degenerate fermion EOS \citep{bisnovatyi2011stellarvol2}, provides another key example.  However, while polytropes with small indices $0<n<1/2$ offer useful crude approximations for planetary objects, they inherently assume a zero density at zero pressure, which is unrealistic for solid and liquid bodies. This limitation restricts their use in planetary science, especially for exoplanets and small Solar System bodies, where non-zero densities are fundamental. Consequently, researchers often face a stark choice: either employ the overly simplistic constant-density model or resort to complex analytical formulations like the Birch-Murnaghan EOS \citep{PhysRev.71.809, doi:10.1073/pnas.30.9.244} or numerical solutions with tabulated data \citep{Smith2018}.

To address this challenge, we introduce a simple, analytical model for perfectly elastic, self-gravitating bodies, characterized by an uncompressed density $\rho_0$ and constant bulk modulus $K$. This leads to a logarithmic EOS that is only approximate, valid for small compression ratios, but is treated as exact within the framework of our study. The proposed model provides a crucial bridge between the overly simplistic constant-density approximation and the complexities of realistic solid-state EOSs, enabling a smooth transition between these two regimes.

EOS mathematically similar to ours have been studied under the name ''logotropic'' model in the context of alternative cosmology and Dark Matter halos by \cite{Chavanis2015} in a series of articles \citep{CHAVANIS2007140,Chavanis2015,  Chavanis_2015,   CHAVANIS2019100271, CHAVANIS201659, PhysRevD.106.063525,Chavanis_2017}. The term ''logotropic'' itself was introduced by 
\citet{MP96,McLaughlin_1997} 
in their study of giant molecular clouds (GMCs). Although the term 'logotropic' ('logatropic') originates in astronomy and cosmology \citep{MP96,Chavanis2015}, here it describes elastic matter with a fixed bulk modulus, distinct from ISM or dark matter models. These previous studies, like ours, feature the density logarithm $\ln{\rho}$, but our work is distinguished by its specific EOS formula, physical scales, and, crucially, interpretation. Previous authors did not recognize the fact that logarithmic EOS  can arise naturally from the assumption of perfectly elastic matter with non-zero density and fixed bulk modulus -- a common characteristic of materials like metals and liquids. Such an approximation is plausible for small compressions and 'cold EOS' conditions where thermal pressure is not dominant. For technical simplicity, our model also omits phase transitions, which may result in abrupt changes of EOS parameters, here treated as constants.

Additional motivation comes from the recent detection of interstellar objects like 1I/'Oumuamua \citep{Meech2017,2018Natur.559..223M} and 2I/Borisov \citep{2019ApJ...886L..29J,Guzik2020} entering our Solar System. They provide direct samples of material formed around other stars. Their sometimes puzzling characteristics, along with the highly anomalous composition of presolar materials found within meteorites, such as the Hypatia stone \citep{KRAMERS2022115043}, suggest that the composition and physical properties of extrasolar planetesimals may differ significantly from those familiar within our own system. Rapidly assessing the nature and structure of future discoveries, potentially composed of unexpected materials, necessitates simple and flexible physical models. The elastic model presented here, relying only on the fundamental parameters of uncompressed density ($\rho_0$)  and bulk modulus ($K$), offers such a tool. It allows for straightforward analytical estimates of internal structure, stability limits, and the mass-radius relation across a vast parameter space, providing crucial first-order insights when detailed equations of state are unavailable or uncertain.

The article is organized as follows. Section \ref{sec:model} presents the derivation of the elastic model and the conversion of the equilibrium equations into a dimensionless form with appropriate scaling.  Section \ref{sec:math} details the properties of the special ''elastic function''  $U(x)$ including its mathematical characteristics, approximations, and asymptotic behavior.
Physical quantities expressed using appropriate scaling and elastic function $U(x)$ are presented in Sect.~\ref{sec:properties}, along with illustrative examples.
Section \ref{sec:astro} explores the astrophysical properties of the model, focusing on the mass-radius $M(R)$ curve, the existence of a maximum mass, and stability considerations. A collection of useful simple approximate analytical formulas is provided in Sect.~\ref{sec:universal_profile}. The limits of applicability of the model are established in Section \ref{sec:realEOS} through a comparison with a realistic iron equation of state.  Finally, Section \ref{sec:conclusions} summarizes the key findings. The appendix provides a detailed, simultaneous derivation of the equations for several analytical equations of state commonly used in astronomy and planetary science.

\section{Elastic Model} \label{sec:model}

We consider a perfectly elastic body with bulk modulus $K$ compressed due to its own gravity. With spherical symmetry, other elastic properties of the matter can be neglected. We also assume uncompressed density\footnote{In \citep{Chavanis2015} $\rho_0$ is of the order of the Planck's density.} $\rho_0$. For example, one could imagine an iron ball with progressively increasing mass (and radius). Initially, it will grow in size at nearly constant density, transitioning from an asteroid to a planetoid, moon/planet, super-Earth, etc. Eventually,  gravity will compress the matter. The subsequent behavior,  and whether this process can continue indefinitely within Newtonian gravity and assuming perfectly elastic matter, will be determined from the analysis of our model.

From the definition of the bulk modulus measured at~$\rho_0$
\begin{equation}
\label{bulk_modulus}
K = \rho \frac{dP}{d\rho}
\end{equation}
we obtain a perfectly elastic equation of state (EOS) in logarithmic form
\begin{equation}
\label{ElasticEOS}
P(\rho) = K \ln{(\rho/\rho_0)}.
\end{equation}
This EOS is valid for $\rho \geq \rho_0$, ensuring $P \geq 0$. The above equation belongs to the barotropic class, along with other important examples such as polytropic and degenerate electron gas EOSs. In non-standard cosmology \citep{Chavanis2015} and interstellar medium research \citep{McLaughlin_1997}  the formula \eqref{ElasticEOS} is known as the logotropic EOS. Using the dimensionless compression ratio (or simply \textit{compression})
\begin{equation}
\label{compression_ratio}
    \eta = \frac{\rho}{\rho_0}
\end{equation}
the elastic EOS can be written as $P = K \ln{\eta}$. 

In engineering, the term  \textit{dilatation}, denoted by
\begin{equation}
\label{dilatation}
    \mu = \frac{\rho}{\rho_0}-1   = \eta - 1,  
\end{equation}
is often used. This casts the elastic EOS in the form  $P = K \ln{(1+\mu)}$. 
 The elastic approximation is valid when the dilatation is small, i.e., when $|\mu| \ll 1$.

We will start with the hydrostatic equilibrium equation in the form
\begin{equation}
\label{eq:hydro}
    h + \Phi_g = C,
\end{equation}
where $h = \int dp/\rho$ is specific enthalpy and $C$ is an integration constant. Gravitational potential $\Phi_g$ obeys Poisson's equation  
$$
\Delta \Phi_g = 4 \pi G \rho.
$$

For EOS \eqref{ElasticEOS} the specific enthalpy is
\begin{equation}
\label{ElasticEnthalpy}
    h(\rho) = -\frac{K}{\rho} + const.
\end{equation}
If we demand that $h(\rho)=0$ at the surface, the constant in \eqref{ElasticEnthalpy} is equal to $K/\rho_0$. Taking spherical 3D Laplacian of \eqref{eq:hydro} yields
\begin{equation}
\label{eq:LE-spherical}
r^{-2} \partial_r \left (r^2 \partial_r h \right ) + 4 \pi G \rho = 0.
\end{equation}

Using a more convenient form of the enthalpy,
\begin{equation}
\label{eq:state}
h = K \int \frac{d\rho}{\rho^2}  = K \int \frac{\rho'}{\rho^2} dr,
\end{equation}
from eqns.~(\ref{eq:LE-spherical})~and~(\ref{eq:state}) we obtain
\begin{equation}
\rho'' + \frac{2\rho'}{r} - \frac{2(\rho')^2}{\rho} + \frac{4\pi G}{K} \rho^3 = 0.
\end{equation}

Substituting $\rho(r) \to \rho_c u(r)$ and dividing through by $\rho_c$ gives a differential equation in $u(r)$:
\begin{equation}
u'' + \frac{2u'}{r} - \frac{2(u')^2}{u} + \frac{4\pi G \rho_c^2}{K} u^3 = 0.
\end{equation}
Using the dimensionless variable  $x = r/\lambda$, 
where the scale length  $\lambda$ is defined as
\begin{equation}
\label{eq:lambda}
    \lambda^2 = \frac{K}{4 \pi G \rho_c^2},
\end{equation}
we remove physical constants and arrive at a crucial ordinary differential equation (Elastic ODE)
\begin{equation}
    \label{elastic_equation}
    u'' + \frac{2}{x} u' - \frac{2 u'^2}{u} + u^3=0.
\end{equation}
This form isolates the nonlinear elastic term $-2 u'^2/u$, distinguishing it from the Lane-Emden equation.
Without this negative term, \eqref{elastic_equation} would be the $n=3$ Lane-Emden equation. The first two terms represent the spherical Laplacian in variable $x$, allowing us to write Eqn.~\eqref{elastic_equation} as
$$
\Delta u - \frac{2 u'^2}{u} + u^3=0,
$$ 
or, even more simply, noting that the reciprocal appears in the expression for the specific enthalpy (Eq.~\ref{ElasticEnthalpy}), we can rewrite the equation as:
\begin{equation}
\label{eq:Delta_inv_u}
\Delta \left( \frac{1}{u} \right) - u = 0.
\end{equation}

The standard initial conditions for \eqref{elastic_equation} are
\begin{equation}
\label{initial_conditions}
    u(0)=1, \quad u'(0)=0.    
\end{equation}

Similar analysis can be performed for the continuity equation 
$$
\frac{d m}{dr} = 4 \pi r^2 \rho.
$$
Substituting
$$
m(r) = 4 \pi \lambda^3 \rho_c \, \mu(r/\lambda)
$$
leads to the dimensionless form
$$
\mu' = x^2 u.
$$
Moreover, due to Eqn.~\eqref{eq:Delta_inv_u}, or by integrating the continuity equation, we obtain the first integral
\begin{equation}
\label{eq:first_integral}
\mu u^2 + u' x^2 = 0.
\end{equation}

\begin{figure}
\includegraphics[width=\linewidth]{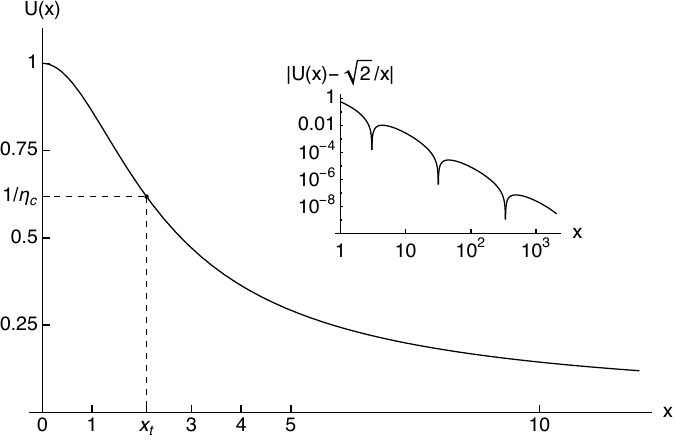}
\caption{\label{fig:U_x} Universal elastic (logotropic) function $U(x)$.  Cutoff value $1/\eta_c=\rho_0/\rho_c$ defines surface of the body and corresponding  dimensionless radius $x_t$. Important one-to-one property is $U(x_t)=1/\eta_c$. Inset illustrates oscillatory asymptotic, which is too small to be visible in the main plot.}
\end{figure}

\section{Mathematical Properties of the Elastic Function \label{sec:math}}

Because all physical properties of our model are described by a single special function, it is worthwhile to study it in detail. This function is defined as the specific solution to Eqn.~\eqref{elastic_equation} with the initial conditions given by \eqref{initial_conditions}. To avoid confusion with the general solution, $u(x)$, we denote this specific solution as the \textit{elastic function}, using the capital letter $U(x)$.

The initial conditions \eqref{initial_conditions} cannot be applied directly at $x=0$ due to the division by zero in the spherical Laplacian. The standard procedure in such a case is to begin integration with a series expansion:
\begin{equation}
\label{eq:szereg}
    U(x) \simeq 1 - \frac{1}{6} x^2 +  \frac{13}{360} x^4 - \frac{25}{3024} x^6 + \ldots
\end{equation}
Comparing this series with that of the Lane-Emden function, we find that the best match is for the polytropic index $n=13/3\simeq4.33$. However, starting with the $x^6$ term, the elastic and Lane-Emden functions diverge. Like for the Lane-Emden equation \citep{Horedt2004}, the series expansion, Eqn.~\eqref{eq:szereg}, has a small radius of convergence, therefore providing a useful solution only for small $x$, but it is not sufficient to cover the entire domain of interest. Nevertheless, the series expansion is important for initiating the numerical integration of Eqn.~\eqref{elastic_equation}, and for deriving a variety of approximate analytical formulas (Sect.~\ref{sec:universal_profile}) such as for the density profile.

\begin{figure}
\includegraphics[width=\linewidth]{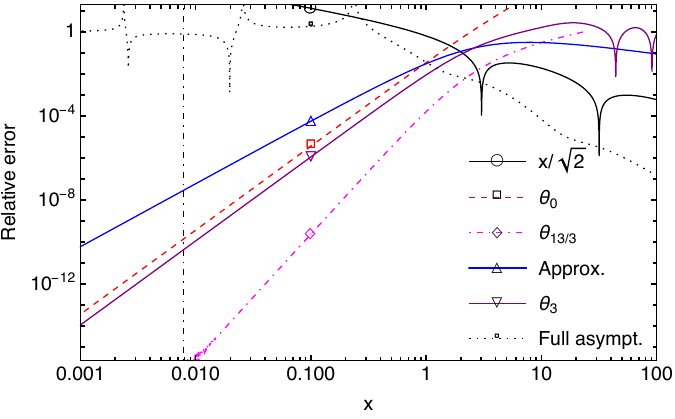}
\caption{\label{fig:Elastic_function_approx} Numerical solution to \eqref{elastic_equation} and various attempts to approximate it using known formulas. Vertical grid line shows $\epsilon = 1/128$ where initial conditions for numerical integration 
were imposed using 6th order series expansion \eqref{eq:szereg}. Curves display absolute value of the error relative to $U(x)$ for simple asymptotics (black, solid), 2nd order series expansion at zero (red, dashed), $n=13/3$ Lane-Emden function $\theta_{13/3}$ (magenta, dot-dashed), analytical formula \eqref{eq:integral_von_neumann} preserving both 2nd series at zero and simple asymptotics (blue, solid),  $n=3$ Lane-Emden $\theta_3$ (purple, solid) and full asymptotic formula \eqref{oscillatory_asymptotics} (black, dotted).
}
\end{figure}

The elastic function $U(x)$, obtained as a numerical solution to the ODE, is presented in Fig.~\ref{fig:U_x}. From numerical evidence, we infer that $U(x)>0$, i.e., the solution is positive everywhere (see also \citealt{Chavanis2015}).
The solution asymptotes to 
\begin{equation}
\label{eq:asymptotic_formula}
    U(x) \sim \frac{\sqrt{2}}{x},
\end{equation}
which is also the exact singular solution of Eqn.~\eqref{elastic_equation}. Therefore, in analogy to the n=5 Plummer solution, $\theta_5(x) = 1/\sqrt{1+x^2/3}$ \citep{Mach5}, we attempted to find a simple analytical formula preserving both Eqns.~\eqref{eq:szereg} and \eqref{eq:asymptotic_formula}. The best result, obtained 
using the integral form, Eqn.~\eqref{eq:integral_eqn}, reads
\begin{equation}
    \label{eq:integral_von_neumann}
    U \simeq \frac{1}{ 3 + 1/y - 2(1+y) \ln{(1+1/y)}}, \; y=\frac{\sqrt{2}}{x}.
\end{equation}
See Fig.~\ref{fig:Elastic_function_approx} to assess the quality of this simple formula, for which error relative to $U(x)$ is plotted as a solid blue line. As we will demonstrate, such a simple formula is not generally applicable, and the behavior of the model depends strongly on the actual solution in the range not covered by either the series or asymptotic formulas. A~high-precision numerical solution to Eqn.~\eqref{elastic_equation} for the elastic function is used as a reference in Fig.~\ref{fig:Elastic_function_approx}. Possible approximations include: the Lane-Emden function with $n=3$ (Fig.~\ref{fig:Elastic_function_approx}, purple), $n=13/3$ (Fig.~\ref{fig:Elastic_function_approx}, magenta, dot-dashed) and the asymptotic formula (Fig.~\ref{fig:Elastic_function_approx}, black).  These approximations are summarized in Fig.~\ref{fig:Elastic_function_approx}. For completeness, we recall that the $n=0$ Lane-Emden function, $\theta_0(x) = 1-x^2/6$ (Fig.~\ref{fig:Elastic_function_approx}, red, dashed),  corresponds to a constant density $\rho(r)=\rho_0$ ball in polytropic\footnote{This is because, in the polytropic model, the density is given by $\rho=\rho_c \theta_n^n$. Therefore, for
$n=0$, we raise the Lane-Emden function to the zero-th power, resulting in a constant density.}, but not in the elastic model. For the latter, it is still a well-behaved approximation for $U(x)$.

Function $U$ is monotonic, so it makes sense to define its inverse $U^{-1}$, which is uniquely defined for arguments in the interval $(0,1)$. Although conceptually useful, the inverse function can sometimes be challenging to work with in practice; therefore, we primarily rely on the parameters $\eta_c$ and $x_t$ (see Fig.~\ref{fig:U_x}) in subsequent derivations.

Elastic equation, Eqn.~\eqref{elastic_equation}, has many interesting properties. If we introduce the reciprocal function $w=1/u$, a particularly appealing form of it is obtained:
\begin{equation}
\label{logotropic_eqn}
    \Delta w = \frac{1}{w}.
\end{equation}
This can be considered as a Lane-Emden \eqref{Lane-Emden} equation with index $n=-1$ \citep{McLaughlin_1997,PhysRevD.106.063525}. \cite{KYCIA20151003} classify it as a generalized Emden–Fowler equation with parameters, in their notation: $\alpha=2,\delta=-1, n=0$ and $p=-1$.
Despite its simplicity and many applications in astrophysics \citep{MP96, Chavanis_2015}, the logotropic equation, Eqn.~\eqref{logotropic_eqn}, has not been studied in the detail required for this work. The simpler form of \eqref{logotropic_eqn} compared to \eqref{elastic_equation} is not surprising, given the function $w$ being a linear function of the specific enthalpy, Eqn.~\eqref{ElasticEnthalpy}, which is known to be the optimal function for
computations related to barotropic configurations \citep{1985A&A...146..260E,10.1046/j.1365-8711.2003.06951.x,Odrzywolek_PhD}.

Initial conditions for $w$, $w(0)=1/u(0)=1, w'(0)=- (1 / u(0)^2) u'(0) = 0$, are identical to \eqref{initial_conditions}. A formal series solution for $w$ is
$$
w(x)  = \sum_{n=0}^\infty \frac{c_n}{(2n+1)!} x^{2n},
$$
where the coefficients obey the recurrence relation:
$$
c_n = \frac{1}{2n^2+n}\sum_{k=1}^{n-1} \binom{2 n+1}{2 k+1} c_k (k-n) c_{n-k}.
$$
The first coefficients are $c_0=1,c_1=1,c_2 = -1, c_3 = 13/3, c_4 = -125/3$, \ldots
The radius of convergence can be estimated numerically from
$$
\lim_{n\to \infty} \frac{c_{n}}{c_{n+1}} (2n+3)(2n+2) \simeq 4.
$$

The function
$$w(x)=x/\sqrt{2},$$
is an exact analytical solution to Eqn.~\eqref{logotropic_eqn}, 
which, remarkably, also describes the asymptotic behavior of solutions to this equation as $x \to \infty$. However, the above asymptotics provide only a crude approximation for large $x$. In the next section, we will need more terms, which can be found as follows. First, we substitute
$$
w(x) = \frac{x}{\sqrt{2}} + \epsilon \, \phi(x),
$$
into \eqref{logotropic_eqn} where $\epsilon$ is a small perturbative parameter. A series expansion with respect to $\epsilon=0$ gives:
$$
\left( \phi''  + \frac{2 \phi'}{x}  + \frac{2\phi}{x^2}  \right) \epsilon - \frac{2 \sqrt{2}}{x^3} \phi^2 \epsilon^2 + O(\epsilon)^3 = 0,
$$
where the zeroth-order term in $\epsilon$ vanishes. Ignoring the secular $\epsilon^2$ term, we obtain a linear ODE: 
$$
\phi''  + \frac{2}{x} \phi'  + \frac{2}{x^2} \phi = 0,
$$
with the analytical solution
\begin{equation}
\label{oscillatory_asymptotics}
\phi(x) = A \cos{\left( \frac{\sqrt{7}}{2} \ln{(x)} + \phi_0 \right)/\sqrt{x}}.
\end{equation}
 The coefficient $A \simeq 0.4164$ and phase $\phi_0 \simeq 0.1395$ can be determined numerically to match the numerical solution (Fig.~\ref{fig:Elastic_function_approx}, black dotted curve). Attempts to find these values analytically using textbook methods \citep{Olver1974} were unsuccessful. This subtle oscillation described by Eqn.~\eqref{oscillatory_asymptotics} is numerically so small that it is barely visible on standard plots (cf.~Fig.~\ref{fig:U_x}, inset, or Fig.~18 in \citealt{Chavanis2015}). Nevertheless, it plays a crucial role in explaining the unexpected spiraling behavior observed in the mass-radius relation (Fig.~\ref{fig:MR_combined}), discussed further in Sect.~\ref{sec:astro}. Due to its subtlety and unexpected nature, we explored its origin more deeply.

The oscillatory behavior, which at first seems to emerge out of nowhere, can be understood in the following way. Changing the independent variable in \eqref{logotropic_eqn} to $x=e^t$ we obtain:
$$
W''+W' = \frac{e^{2t}}{W},
$$
where $W(t) \equiv w(e^t)$. Rescaling $W(t) = e^t z(t)$ leads to the autonomous system
$$
z''+3 z' + 2z-1/z = 0.
$$ 
This equation can be cast in an easily recognizable  energy-conservation form from classical mechanics:
$$
\begin{cases}
    \frac{d}{dt} \left[ \frac{1}{2} v^2 + V(z) \right] = -\kappa v^2\\
    \frac{dz}{dt} = v
\end{cases}
$$
where the potential $V$ is
$$
V(z)  = z^2 - \ln{z}.
$$
Therefore, the solution of the logotropic equation \eqref{logotropic_eqn} is equivalent to classical one-dimensional motion in a potential well, $V(z)$, with energy dissipation, resembling a typical damped oscillator system. The quadratic drag coefficient, $\kappa=3$, is very close to the threshold of 4, where oscillations vanish in the linear regime. This explains why oscillations were difficult to detect in the numerical results. The specific solution required for our problem is the one with infinite initial energy, falling down the potential well until the minimum, $V_\text{min} = \ln{\sqrt{2e}}$, is reached at $z=1/\sqrt{2}$. The trajectory equivalent to function $U$ on $\{z,v\}$ plane is the one starting from the straight line $v=-z$.

We also note the Volterra integral equation equivalent to our elastic problem
\begin{equation}
\label{eq:integral_eqn}
w(x) = 1+ \int_0^x \left( s - \frac{s^2}{x} \right) \frac{ds}{w(s)},
\end{equation}
which can be used to find iteratively a variety of approximations, including Eqn.~\eqref{eq:integral_von_neumann}.

\section{Physical Properties of the Elastic Model \label{sec:properties}}

Using $U(x)$ and its inverse, we can derive formulas for all the interesting properties of the self-gravitating elastic body. We denote the scaling radius as
\begin{equation}
\label{R0}
    R_0^2 = \frac{K}{4 \pi G \rho_0^2}.
\end{equation}
Using sound speed, $c_s$, at $\rho_0$
\begin{equation}
\label{sound_speed}
c_s^2 = \frac{\partial P}{\partial \rho} \Bigg|_{\rho_0} = \frac{K}{\rho_0},
\end{equation}
we can express the radial scale in terms of the sound crossing time:
$
R_0 = 1/(\pi \sqrt{3/8}) c_s \tau \simeq 0.52 c_s \tau,
$
where $\tau = \sqrt{3\pi/(32 G \rho_0)}$ is the dynamical free-fall collapse timescale of a uniform ball, assuming zero pressure. 

Note that comparing Eqns.~\eqref{eq:lambda} and \eqref{R0}, we get $R_0 = \lambda \eta_c$, where  
the dimensionless central compression ratio is
\begin{equation}
\label{eq:eta_c}
    \eta_c = \frac{\rho_c}{\rho_0}.
\end{equation} 
The related dimensionless radius, $x_t$, is connected to $\eta_c$ (Figure~\ref{fig:U_x}) via the elastic function, $U(x)$. The essential relation
\begin{equation}
\label{eq:eta_c_inv}
    \frac{1}{\eta_c} = U(x_t),
\end{equation} 
allows us to use $\eta_c$ or $x_t$ interchangeably. Physically, this relation explicitly defines the surface of the body as the radius where density reaches the uncompressed density $\rho_0$.
The mass scale is simply
\begin{equation}
\label{M0}
    M_0 = \frac{4}{3} \pi \rho_0 R_0^3.
\end{equation}

Example values of the uncompressed density $\rho_0$, bulk modulus $K$, and the resulting scaling parameters, $R_0$ and $M_0$, for selected materials are given in Table~\ref{tbl:K_rho_tbl}.
For liquids, the bulk modulus is directly proportional to the sound speed squared $c_s^2$, \emph{via} Newton-Laplace equation, eq.~\eqref{sound_speed}. For isotropic solids, $K$ can be derived from any two other elastic moduli, e.g., $K=\tfrac{1}{3} E/(1-2 \nu)$, where $E$ is the Young modulus and $\nu$ - Poisson ratio. Alternatively, using sound speeds, we recall that longitudinal $c_L$ and shear $c_S$ wave speeds are not identical and the proper formula is
$$
K = \rho_0 \left( c_L^2 - \frac{4}{3} c_S^2 \right).
$$

The materials selected for Table~\ref{tbl:K_rho_tbl} illustrate the wide range of parameters relevant to astrophysical bodies and engineering applications, as well as theoretical extremes. Typical Solar System (SS) materials like iron, silicates (quartz), common ices (water, nitrogen ice), and liquids (liquid nitrogen, liquid methane) are included.  The potential relevance of nitrogen ice is highlighted by hypotheses concerning 'Oumuamua \citep{https://doi.org/10.1029/2020JE006807}. Exploring the parameter space further, Table~\ref{tbl:K_rho_tbl} shows elements resulting in extreme scaling values: mercury (Hg, high $\rho_0$, low $K$) and liquid helium (very low $K$) are on the lower-left corner of the $R_0,M_0$ plane. Boron (B, \citealt{2017_Chuvashova}) and diamond (C, \citealt{Hu2021:Diamond}) provide extremes on the opposite corner (large $R_0,M_0$), due to their very large values of $K$ and small density. 
Osmium, the common-sense record-holder for density $\rho_0$, surprisingly also possesses the highest known bulk modulus $K$ of any element \citep{Dubrovinsky2015}. 
Notably, the calculated mass scale $M_0$ for many typical elements and compounds naturally yields values comparable to the mass of Earth's Moon ($M_L$). This makes the lunar mass a convenient and physically relevant unit for expressing these scales, which vary between $10^{-3} \lesssim M_0/M_L \lesssim 56$ (almost 5 orders of magnitude). The scaling radius across the periodic table of elements is within the range $400 \mathrm{km} \lesssim R_0 \lesssim 7000 \mathrm{km}$
(a factor of $\sim$17). We also provided an amusing, illustrative example of marshmallow. 
Last but not least, strange quark matter is speculated \citep{Kutschera_2020} to exist in astrophysical environments. For the MIT bag model \citep{PhysRevD.9.3471},
as considered in astrophysical contexts \citep{Witten1984:StrangeMatter},
$K=4B/3$, where $B=60$~MeV fm$^{-3}$ is a typical bag constant value. 
The value of $K$ can be only marginally affected by additional factors, such as the inclusion of a non-zero strange quark mass \citep{PhysRevD.110.083041}.

The presented values are intended to provide insight into the possible parameter ranges, using recent experimental and theoretical references where available, rather than a comprehensive survey, and values may depend on specific conditions (temperature, pressure, phase).

\begin{deluxetable*}{lccccccccc}
\tablewidth{\linewidth}
\tablehead{\colhead{Name} & \colhead{Phase} & \colhead{Symbol} & \colhead{$\rho_0$ [g cm$^{-3}$]} & \colhead{K [GPa]} & \colhead{$K'$ }  & \colhead{$R_0$ [km]} & \colhead{$\frac{M_0}{M_L}$} & \colhead{Why incl.?} & \colhead{Ref.}}
\tablecaption{\label{tbl:K_rho_tbl} 
Selected sample of matter with example values of the uncompressed density $\rho_0$, bulk modulus $K$, and its pressure derivative $K'$.  Resulting scale length \eqref{R0} and mass scale \eqref{M0} are provided. Mass scale is given relative to Moon (lunar) mass, denoted by $M_\mathrm{L}$. Only liquid and solid state of matter is considered, and temperature dependence ignored. Due to assumed barotropic form of the EOS and various possible phases, especially for solids, the given values are only rough indicators of the possible measurement ranges. Selected materials represent astrophysically relevant, theoretically interesting, or extreme cases. Values of $\rho_0$ and $K$ are generally well-measured or inferred from sound-speed experiments, while reliable data for the pressure derivative $K'$ are sparse due to experimental challenges and theoretical uncertainties. Thus, $K'$ values are provided only where trustworthy references exist.}
\startdata
\begin{filecontents*}{Table1.csv}
Name         ,          Phase, Symbol  ,  Density,    K    ,   Kprim,  R    , M      ,  Type                  , Ref
Iron         ,          solid, Fe      ,  7.874  ,  159.3  ,  5.86  ,  1780 , 2.5    ,  Inner SS              , \cite{Huang2022:FeNiEOS}
Magnesium    ,          solid, Mg      ,  1.737  ,    45   ,        ,  4220 , 7.4    ,  Inner SS              , \cite{Kaye_and_Laby}
Silicon      ,          solid, Si      ,  2.329  ,    100  ,        ,  4640 , 13.3   ,  Inner SS              , \cite{Kaye_and_Laby}
Quartz       ,          solid, SiO$_2$ ,  2.65   ,   37.4  ,   6.2  ,  2520 , 2.4    ,  Inner SS              , \cite{Wang2015}
Nitrogen     ,            ice, N$_2$   ,  0.85   ,   2.16  ,        ,  1888 , 0.32   ,  Outer SS              , \cite{YAMASHITA2010972}
Nitrogen     ,         liquid, N$_2$   ,  0.807  ,   0.7   ,        ,  1132 , 0.07   ,  Outer SS              , \cite{CRC}
Water        ,            ice, H$_2$O  ,  0.917  ,    8.8  ,        ,  3530 , 2.3    ,  Outer SS              , \cite{young12}
Methane      ,         liquid, CH$_4$  ,  0.44   ,    0.8  ,        ,  2220 , 0.27   ,  Outer SS              , \cite{CRC}
Diamond      ,          solid, C       ,  3.515  ,    439  ,   3.6  ,  6530 , 56     ,  Max. $M_0$            , \cite{Hu2021:Diamond}
Helium       ,         liquid, He      ,  0.125  , 0.004   ,        ,   550 , 1.2e-3 ,  Min. $M_0$            , \cite{CRC}
Mercury      ,         liquid, Hg      ,  13.534 ,    28   ,        ,   425 , 0.06   ,  Min. $R_0$            , \cite{young12}
Boron        , $\alpha$ solid, B       ,  2.36   ,    225  ,   3.0  ,  6940 , 45     ,  Max. $R_0$            , \cite{2017_Chuvashova}
Osmium       ,          solid, Os      ,  22.59  ,   400   ,   4.0  ,   970 , 1.2    ,  Max. $\rho_0;K$       , \cite{Dubrovinsky2015}
Marshmallow  ,           foam, -       ,  0.36   , 8e-5    ,        ,  27   , 4e-7   ,  Illustrative          , \cite{Kelly2014}
Quark matter ,     speculated, -       ,  4.7e14 , 1.3e25  ,    0?  ,  8.4  , 1.6e7  ,  Exotic/Strange        , \cite{Kutschera_2020}
\end{filecontents*}
\csvreader[separator=comma]{Table1.csv}{Name=\name, Phase=\phase, Symbol=\symbol, Density=\density, K=\K, Kprim = \Kprim, R = \R, M=\M, Type = \Type, Ref = \Ref}
{\name & \phase & \symbol & \density & \K & \Kprim & \R & \M & \Type & \Ref \\}
\enddata
\end{deluxetable*}

The quantities $R_0$ and $M_0$ provide  the characteristic length and mass units (kilometers and masses comparable to the Moon), confirming their suitability for analyzing planetoid-sized objects (Table \ref{tbl:K_rho_tbl}). We denote the total radius and mass of the body as $R$ and $M$, respectively.  The corresponding dimensionless radius and mass will be defined as $\tilde{R} = R/R_0$ and $\tilde{M} = M/M_0$, respectively.

The function $U(x)$, with the appropriate length scale \eqref{eq:lambda}, leads to formulas for the physical description of the self-gravitating body, in analogy to the Lane-Emden and Chandrasekhar models (Appendix~\ref{ThreeModels}). For the density stratification, we have
\begin{equation}
\label{eq:rho_profile}
    \rho(r) = \rho_c \, U(r/\lambda).
\end{equation}
The enclosed mass is
\begin{equation}
    m(r) = \kappa \, \mu(r/\lambda), \quad \kappa = 4 \pi \rho_c \lambda^3,
\end{equation}  
where $\mu$ can be expressed using $U$ via Eqn.~\eqref{eq:first_integral}.
Therefore, for the gravitational acceleration $g=Gm/r^2$, we have
\begin{equation}
    g(r) = -4 \pi G \rho_c  \lambda  \; \frac{U'(r/\lambda)}{U(r/\lambda)^2}.
\end{equation}
The maximum g-force is always at the surface of the elastic body because $g(r)$ is a monotonically increasing function.

The uncompressed density, $\rho_0$,
provides a cutoff and defines the physical radius, $R$, and mass, $M$, of the body. 
Using Eq.~\eqref{eq:eta_c_inv} the dimensionless radius of the body is therefore explicitly given as:
\begin{equation}
\label{R}
    \tilde{R} = \frac{x_t}{\eta_c} = x_t U(x_t).
\end{equation}
Correspondingly, using Eqns.~\eqref{eq:eta_c_inv} and \eqref{eq:first_integral}, the dimensionless mass can be succinctly expressed as:
\begin{equation}
\label{M}
    \tilde{M} = -3 \left( \frac{x_t}{\eta_c} \right)^2 \frac{U'(x_t)}{U(x_t)^2} = -3 x_t^2 U'(x_t).
\end{equation}

The density contrast is
\begin{equation}
\label{density_contrast}
    \frac{\rho_c}{\bar{\rho}} = -\frac{x_t}{3} \frac{U(x_t)^2}{U'(x_t)},
\end{equation}
where the mean density $\bar{\rho} = M/(\frac{4}{3} \pi R^3)$. The latter can be expressed in units of $\rho_0$:
\begin{equation}
\label{avg_density}
\frac{\bar{\rho}}{\rho_0} = - \frac{3 U'(x_t)}{x_t U(x_t)^3}.
\end{equation}

The gravitational potential at the center is given by:
$$
\Phi_g(0) = -\frac{GM}{R}
\left\{  
1 + \frac{ [ U(x_t)-1 ] U(x_t)}{x_t U'(x_t)} 
\right\}.
$$
Coefficient in curly braces increases from the well-known result of $3/2$ for a uniform ball to $2$ as $x_t \to \infty$.

The gravitational binging energy $\Omega$ is:
\begin{equation}
    \Omega = - \frac{G M^2}{2 R} \left \{   1 - \frac{U(x_t)}{x_t U'(x_t)} - \frac{U(x_t)^4}{3 U'(x_t)^2} \right \}.
\end{equation}
To calculate the limit of the coefficient in curly braces for $x_t \to 0$, which is, as expected from a well-known result for uniform ball, 6/5, at least the 4-th order expansion \eqref{eq:szereg} of $U$ is required. For $x_t \to \infty$, using \eqref{eq:asymptotic_formula}, we obtain 4/3.

\section{Mass-Radius Relation and Stability of Elastic Bodies \label{sec:astro} }

\begin{figure*}
\includegraphics[width=\linewidth]{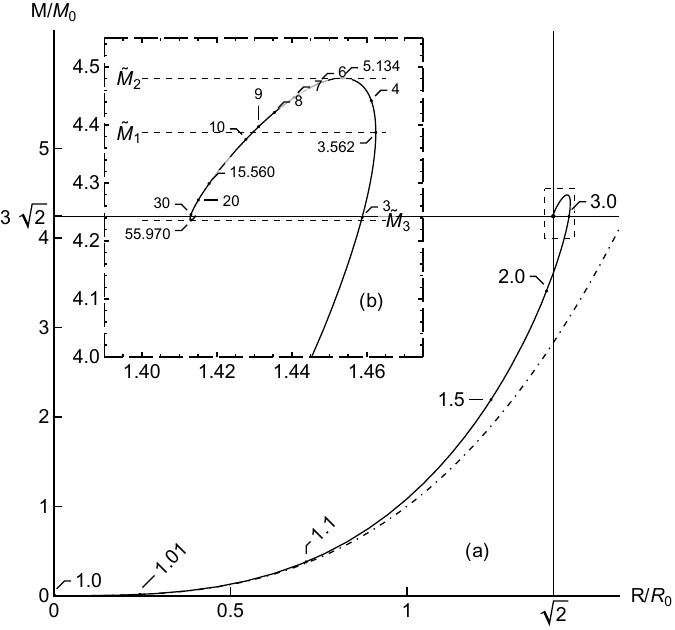}
\caption{
\label{fig:MR_combined}
Mass-radius relation for elastic model: (a) full curve showing the relationship between mass and radius, with grid line crossing marking the asymptotic point; (b) zoomed view of the unstable branch.}
\end{figure*}

A crucial tool in the analysis of self-gravitating bodies with a given EOS is the mass-radius relation, $M(R)$. For the elastic model, it is given in parametric form by equations \eqref{R} and \eqref{M} and is plotted in Fig.~\ref{fig:MR_combined}. The callout points on the mass-radius curve in Fig.~\ref{fig:MR_combined} indicate the values of the central overdensity parameter, $\eta_c$. As a sanity check, we note that for a body of constant density, the dimensionless relation is simply
$$
\tilde{M}= \tilde{R}^3.
$$

In the main panel (a) of Fig.~\ref{fig:MR_combined}  the above relation (a cubic function)  is shown as a dot-dashed line. The constant density approximation is visibly inaccurate even at 10\% compression of the initial density. Surprisingly, the elastic $M(R)$ curve (Fig.~\ref{fig:MR_combined}, (a), solid line) does not extend indefinitely, as one might expect from the constant density approximation, but terminates.

From the asymptotic formula \eqref{eq:asymptotic_formula}, using (\ref{R}, \ref{M}) we can easily find the terminating point, which is achieved in the limit of the central compression parameter $\eta_c\nobreak\to\nobreak\infty$: 
$$
\tilde{R}_\infty = \sqrt{2}, \quad \tilde{M}_\infty = 3 \sqrt{2}.
$$
In other words, in the asymptotic case, the mass-radius relation degenerates to a single point, shown as the intersection of grid lines in Fig.~\ref{fig:MR_combined}~(a).

The behavior near the tip of the $M(R)$ curve is non-trivial. Before the terminating point, we first reach the maximum radius (Fig.~\ref{fig:MR_combined}, inset panel) at
$$
\tilde{R_1} \simeq 1.462, \quad \tilde{M_1} = 4.387.
$$
This is attained at $x_t\simeq 5.21$, which is equivalent to $\eta_c \simeq 3.562$. Not far from that point, at $\eta_c\simeq 5.134, x_t\simeq 7.46$, we achieve the maximum mass:
$$
\tilde{R_2} \simeq 1.453, \quad \tilde{M_2} = 4.481,
$$
cf. panel (b) of Fig.~\ref{fig:MR_combined}. 

The $M(R)$ curve then begins to spiral towards the terminating point, forming a sequence (possibly infinite) of alternating $R$ and $M$ extrema. This conclusion is supported by the asymptotic behavior described in equation \eqref{oscillatory_asymptotics}, and the distance from the terminating point decreases proportionally to $\eta_c^{-3/2}$.

The shape of the $M(R)$ curve, Fig.~\ref{fig:MR_combined} (b), allows us to discuss some aspects of stability for the elastic model \citep[see Sect. 12.3.1]{bisnovatyi2011stellarvol2}. The \emph{turning point theorem} (e.g., \citealt{Zeldovich1963,harrison1965gravitation,H_2021}) indicates that the first mass peak marks the onset of dynamical instability.
A body with mass above the maximum cannot exist in hydrostatic equilibrium and is, therefore, dynamically unstable. For the branch where the solution for a given mass is not unique (see Fig.~\ref{fig:MR_combined}b), secular instability occurs. A process dissipating the binding energy gap can bring the body to a state of smaller radius (the ground state). This is possible for masses above the second extremum (1st local minimum) of mass, at $\eta_c\simeq55.97, x_t\simeq79.1$, where we have
$$
\tilde{R_3} \simeq 1.413, \quad \tilde{M_3} = 4.236.
$$
The secular instability limit is shown as the horizontal grid line labeled $\tilde{M}_3$ in Fig.~\ref{fig:MR_combined}b, just below the line denoting the mass of the terminating point at $\tilde{M}=3\sqrt{2}$.

 Remarkably, the instability  leading to a maximum mass ($M_2$)  is of purely Newtonian nature. To confirm this, we compare the Schwarzschild black hole radius $R_g$ associated with the characteristic mass scale $M_0$, to the model's characteristic radius $R_0$:
$$
R_g/R_0 = \frac{2 G M_0/c^2}{R_0} = \tfrac{2}{3} K/\rho_0/c^2 = \tfrac{2}{3} (c_s/c)^2,
$$
where $c_s$ is the sound speed \eqref{sound_speed} in the medium and $c$ is the speed of light. Using iron as an example, this ratio is extremely small, $R_g/R_0 \simeq 1.6 \times 10^{-10}$, confirming that general relativistic effects are entirely negligible in determining the structure and stability limits found in this model.

One of the motivations for initiating this study was a question, whether one could manufacture spherical bodies with average density much higher than those found in heavy precious metals (platinum, gold, tungsten, osmium, cf.~Table~\ref{tbl:K_rho_tbl}). Such dense, spherical bodies could be then used \citep{1986AJ.....92..986H,Feldman_2016} in deep space experiments \citep{2019PhRvD.100f2003A} to measure the gravitational constant, $G$ \citep{10.1063/1.4994619}. Naively, without our analysis, for large $\rho_0$ and small $K$ one would expect significant compression under self-gravity. The elastic model can answer this question precisely. Using Eqn.~\eqref{avg_density}, the maximum average density of $1.508 \rho_0$ is achieved for $x_t\simeq22.1$ equivalent to $\eta_c \simeq 15.6$. This is deep into the unstable branch of the $M(R)$ curve, cf. inset (b) of Fig.~\ref{fig:MR_combined}. Therefore, in practice, the maximum $\bar{\rho} \simeq 1.46 \rho_0$ is found for maximum mass threshold $M_2$. The question is therefore answered negatively within this model: stable configurations cannot achieve an average density more than about 50\% higher than the initial density $\rho_0$, regardless of how small the bulk modulus is assumed to be.

In summary, the elastic body exhibits secular instability when its dimensionless mass exceeds a critical threshold, specifically $\tilde{M} > \tilde{M_3} \approx 4.236$. Furthermore, it transitions to a state of dynamic instability when the dimensionless mass surpasses a higher critical value, namely $\tilde{M} > \tilde{M_2} \approx 4.481$. These findings, illustrated by the spiraling behavior of the mass-radius curve in Fig.~\ref{fig:MR_combined}b,  encapsulate the primary conclusions drawn from the preceding detailed analysis.

\section{Universal Quadratic Profile \label{sec:universal_profile} }


In Section \ref{sec:math}, we announced various analytical approximations (Fig.~\ref{fig:Elastic_function_approx}) to the elastic function $U(x)$ with the goal of deriving simple analytical formulas. Using  the asymptotic formula, Eq.~\eqref{eq:asymptotic_formula}, yields only the trivial result $\tilde{R} = \tilde{R}_\infty = \sqrt{2}$, $\tilde{M} = \tilde{M}_\infty = 3 \sqrt{2}$, i.e., the mass-radius curve degenerates to a single point.

However,  a more useful approximation is obtained by replacing the elastic function, $U(x)$, with the $n=0$ Lane-Emden function, $\theta_0(x)$, which is equivalent to the series expansion \eqref{eq:szereg} truncated at the quadratic term. Substituting $U(x) \to 1-x^2/6$ into equations  \eqref{R} and \eqref{M} leads to
\begin{equation}
\label{MR_universal}
    \tilde{R} = \sqrt{6} \frac{\sqrt{1-1/\eta_c}}{ \eta_c}, \quad \tilde{M} = 6 \sqrt{6} (1-1/ \eta_c)^{3/2}. 
\end{equation}
The formula for \emph{compactness}, 
$$
\frac{ \tilde{M} }{ \tilde{R} } = 6 \mu_c,
$$
is particularly simple.

The corresponding density profile \eqref{eq:rho_profile}, expressed in terms of the radial variable, becomes:
\begin{equation}
\label{quadratic_profile}
    \rho(r) = \rho_c \left( 1 - \frac{r^2}{R^2} \right) + \rho_0 \frac{r^2}{R^2}. 
\end{equation}
From this, the mass is
\begin{equation}
\label{quadratic_profile:M}
M=\frac{4}{3} \pi R^3 \left( \tfrac{2}{5} \rho_c + \tfrac{3}{5} \rho_0 \right) = \frac{4}{3} \pi R^3 \rho_0 (1+ \tfrac{2}{5} \mu_c),
\end{equation}
allowing for a quick estimate of the central density given $M, R$ and $\rho_0$. The moment of inertia is given by the similarly elegant formula:
\begin{equation}
\label{quadratic_profile:I}
I=\frac{8}{15} \pi R^5 \left( \tfrac{2}{7} \rho_c + \tfrac{5}{7} \rho_0 \right)
=
\frac{2}{5} M R^2 \frac{1+ 2 \mu_c/7}{1+ 2 \mu_c/5}. 
\end{equation}
Due to central compression, the moment of inertia for an elastic body is smaller than that of a uniform ball:
$$
I \simeq \frac{2}{5} M R^2 (1 - \tfrac{4}{35} \mu_c + \tfrac{8}{175} \mu_c^2 + \ldots )
$$

We also present formulas for gravitational potential at the center,
\begin{equation}
\label{quadratic_profile:Phi}
\Phi_g(0) = - \pi G (\rho_c + \rho_0) R^2,
\end{equation}
and binding energy, 
\begin{equation}
\label{quadratic_profile:U}
\Omega = -\frac{16 \pi^2 G R^5}{315} \left( 4 \rho_c^2 + 10 \rho_c \rho_0 + 7 \rho_0^2 \right), 
\end{equation}

which might become useful for quick estimates of the energy required to dismantle potentially hazardous near-Earth asteroids, e.g., the recently tracked 2024 YR4 \citep{CNEOS2025}.

Eliminating $\eta_c$ from Eqns.~ \eqref{MR_universal} yields the dimensionless $\tilde{M}(\tilde{R})$ curve, described by a cubic polynomial:
$$
\tilde{M}^3 + 18 \tilde{M}^2 \tilde{R} + \tilde{M} (108 \tilde{R}^3 - 216)  + 216 \tilde{R}^3=0.
$$

A notable property of the simple profile given by Eq.~\eqref{quadratic_profile} is its universality. As we will demonstrate in the next Section \ref{sec:realEOS}, for small overdensities, this profile matches the quadratic expansion common to many physical EOS at low compression. Furthermore, as an approximation, its applicability extends far beyond theoretical expectations. Analogous formulas could be derived for higher-order expansion, but they became too complicated to be practical.

\section{Comparison with Realistic Iron EOS Models  \label{sec:realEOS} }

To compare the elastic model with more realistic cases, we begin with a survey of EOSs. A common starting point for a higher-order analytical description of elastic matter is the Birch–Murnaghan \citep{PhysRev.71.809,doi:10.1073/pnas.30.9.244} EOS 
\begin{equation}
\label{Birch–Murnaghan}
    P(\rho) = \frac{3}{2} K \left( x^7 - x^5 \right) \times \left\{ 1+ \frac{3}{4} (K'-4)\left( x^2-1\right) \right\},
\end{equation}
where 
$$
x = ( \rho/\rho_0)^{1/3}.
$$

The series expansion of the equation \eqref{Birch–Murnaghan} at $\rho=\rho_0$  agrees with \eqref{ElasticEOS} up to quadratic order only when $K'=0$. Dimensionless coefficient $K'$ is
$$
K' = \frac{\partial K}{\partial P}\Bigg|_{P=0}.
$$

At high densities, Eq.~\eqref{Birch–Murnaghan} becomes polytropic with an index $n=1/2$:
$$
P(\rho) \sim \frac{9K}{8} (K'-4) (\rho/\rho_0)^3.
$$
Therefore,  the $n=1/2$ Lane-Emden function is sometimes used to find a crude approximation for Earth-like planetary core. An exceptional case for the Birch-Murnaghan EOS occurs when $K'=4$, where the asymptotic is different:
$$
P(\rho) \sim \frac{3K}{2} (\rho/\rho_0)^{7/3},
$$
i.e., $n=3/4$. Note that the EOS \eqref{Birch–Murnaghan} becomes unphysical for $K'<4$ if densities are larger than
$$
\rho_\text{max} = \rho_0 \left(\frac{-2 \sqrt{9 {K'}^2-84 {K'}+241}+21 {K'}-98}{27(K'-4)}\right)^{3/2},
$$
because pressure starts to decrease with increasing density.

Another option for incorporating stiffening of matter (increasing bulk modulus), especially for small $K'<4$, is to generalize logotropic EOS, in a manner similar to how we derived it from Eqn.~\eqref{bulk_modulus}.  The \textit{generalized} elastic EOS is then obtained by replacing $K \to K + K' P$ in definition \eqref{bulk_modulus} and solving the resulting ODE, producing: 
\begin{equation}
\label{const_Kprim}
    P(\rho) = \frac{K}{K'} \left[ \left( \frac{\rho}{\rho_0} \right)^{K'} -1 \right].
\end{equation}
 The limit of equation \eqref{const_Kprim} as $K'\to0$ \citep{10.1119/10.0000841}, obtained, for example,  using L'Hospital's rule, is equal to formula \eqref{ElasticEOS}.

We remark that in the limit of small central compression $\eta_c \simeq 1$ (dilatation $\mu_c \ll 1$), the elastic EOS is equivalent to the \textit{linear} EOS: 
\begin{equation}
\label{Linear_EOS}
P = K (\rho/\rho_0-1) = K \mu.
\end{equation}
The generalized elastic EOS \eqref{const_Kprim} is also linear when $K'=1$.

If we limit analysis to a specific, well-studied material, like iron, a large set of EOSs for small overdensities is used in engineering -- for example, in crash simulations. Among these, we can list the Compaction EOS (a cubic polynomial in $\mu$), Gruneisen EOS \citep{10.1111/j.1365-246X.1978.tb03757.x}, and the linear EOS \eqref{Linear_EOS}.
Notably, engineering EOSs usually extend below $\rho_0$, due to possible tensile stretching of matter. In astrophysics, this type of behaviour is very unlikely, unless we need to model unusual situations, such as those involving central cavities. The natural occurrences of large spherical central cavities seems unlikely, except in very exotic scenarios, for instance, those involving micro black holes \citep{DAI2024101662}. However, this might be interesting for futuristic scenarios like drilling space habitats or space mining. We do not go into this discussion further and assume $\rho>\rho_0$ for the remainder of this work.

A recent article by \citet{PhysRevB.108.014102} discusses iron EOS over a wide range of temperatures, phases and densities. For our purposes, it is sufficient to use their fit to the cold AP2  model of \citet{Holzapfel_AP2} 
\begin{equation}
\label{AP2_EOS}
P = 3 K \frac{1-x}{x^5} e^{a_0(1-x)} \left( 1+ a_2 x (1-x) \right), \; x = (\rho_0/\rho)^{1/3},
\end{equation}
where $a_0 = - \ln{(3K/P_\text{FGr})}$, $a_2 = (3/2) (K'-3) - a_0$, $P_\text{FGr} = a_\text{FG} (Z/V_0)^{5/3}$. In SI units, $a_\text{FG} = 2.3336 \times 10^{-38}$ [J m$^2$] and $V_0 = m_\text{Fe}/\rho_0$ where the atomic mass for $^{56}$Fe is $m_\text{Fe}=55.934936$~u.

\begin{figure}
    \centering
    \includegraphics[width=\linewidth]{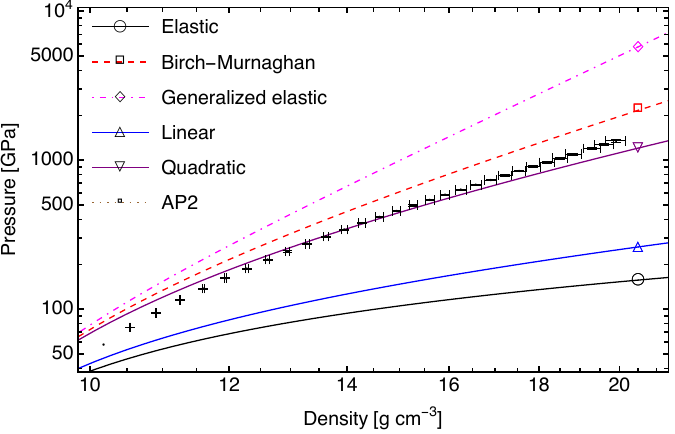}
    \caption{
       \label{fig:ExperimentalEOS_vs_formulas} 
    Comparison of high-density iron EOS measured at National Ignition Facility and various analytical formulas. Data points and errors are from \cite{Smith2018}. Parameters for analytical formulas, from \citet{PhysRevB.108.014102}, Table V, $\alpha$-phase are: $\rho_0 = 7.927$~g~cm$^{-3}$, $K = 164.8$~GPa, $K'= 5.55$. Respective formulas are: Elastic EOS \eqref{ElasticEOS}, Birch-Murnaghan \eqref{Birch–Murnaghan}, Generalized elastic (Power-law) \eqref{const_Kprim}, Linear \eqref{Linear_EOS}, Quadratic \eqref{Quadratic_EOS} and Holzapfel AP2 \eqref{AP2_EOS}.
    }
\end{figure}

State-of-the-art (SOTA) experimental data for high-density $^{56}$Fe EOS have been presented by \cite{Smith2018}. Their Supplementary Table~1 provides both the EOS and the corresponding mass-radius relation. The data starts at compression $\eta \simeq 1.3$ (Fig.~\ref{fig:ExperimentalEOS_vs_formulas}).
Comparing various analytical formulas with numerical data (Fig.~\ref{fig:ExperimentalEOS_vs_formulas}) we found AP2 EOS \eqref{AP2_EOS} performs exceptionally well for high densities, with Birch-Murnaghan \eqref{Birch–Murnaghan} being the next best. Surprisingly, the simple quadratic EOS,
\begin{equation}
\label{Quadratic_EOS}
    P(\mu) = K \mu - \frac{1}{2} K (1-K') \mu^2,
\end{equation}
provides a fit that is almost as good (Fig.~\ref{fig:ExperimentalEOS_vs_formulas}, purple) as these more sophisticated approaches. Notably, equations \eqref{Birch–Murnaghan}, \eqref{const_Kprim} and \eqref{AP2_EOS} all\footnote{This is true also for \eqref{ElasticEOS} for which we have $K'=0$.} have an identical second-order expansion given by \eqref{Quadratic_EOS} and begin to differ only at the third-order expansion in $\mu$ or higher. Therefore, we will use EOS of the form \eqref{Quadratic_EOS} to assess the validity of our elastic model. It is worth noticing that for $\rho<11$~g~cm$^{-3}$, the experimental data (Fig.~\ref{fig:ExperimentalEOS_vs_formulas}, data points with error bars) tend towards the linear and elastic EOS predictions. If this trend is more pronounced for materials other than iron, then the elastic model may,  in fact, have a wider range of applicability than indicated by subsequent analysis.

\begin{figure*}
\gridline{\fig{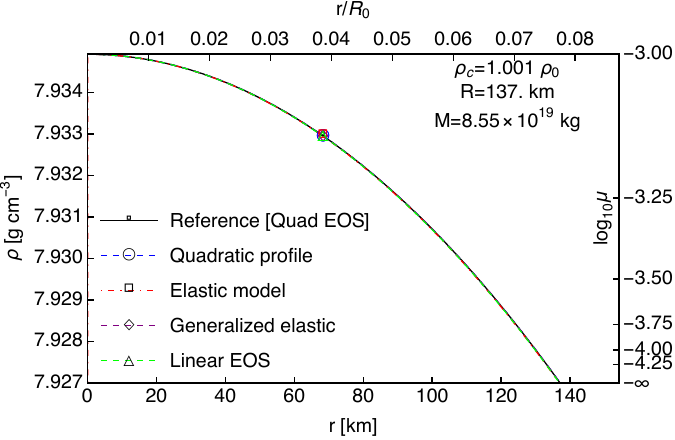}{0.5\linewidth}{(a)}
          \fig{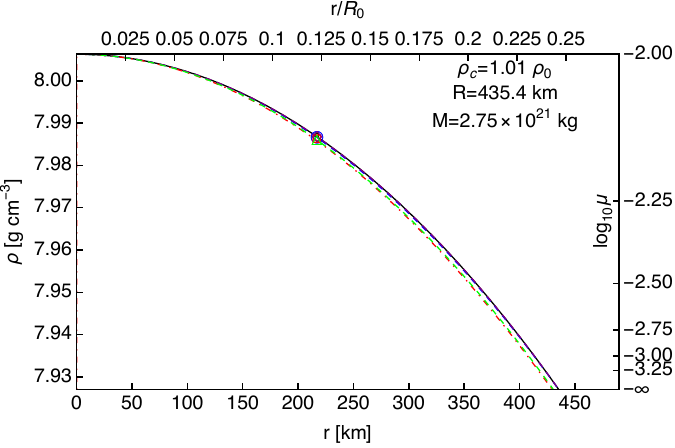}{0.5\linewidth}{(b)}}    
\gridline{\fig{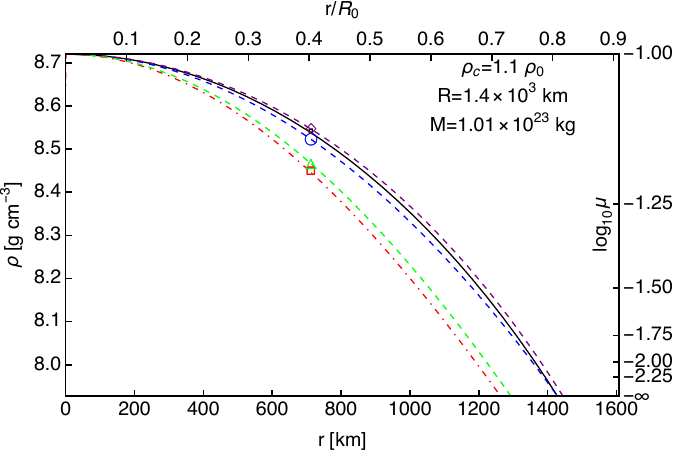}{0.5\linewidth}{(c)}
          \fig{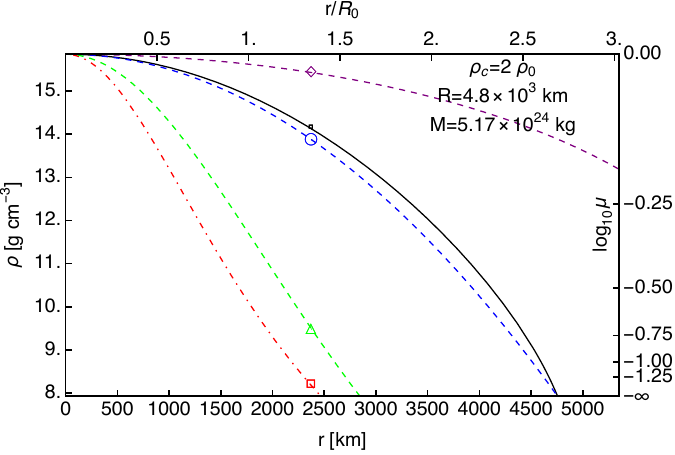}{0.5\linewidth}{(d)}}  

    \caption{Comparison of profile densities. Legend shown in panel (a) is common for all four plots. }
    \label{fig:Comparison}
\end{figure*}

To quantitatively measure the quality of the elastic model approximation, we perform a direct comparison of the internal density profiles (Fig.~\ref{fig:Comparison}). For this, we require a reference model valid from $\eta=1$ upwards. Available high-density experimental data \citep{Smith2018} (Fig.~\ref{fig:ExperimentalEOS_vs_formulas}) exhibits significant error bars and starts at $\eta \approx 1.3$. Constructing a smooth numerical reference across the full range by 'gluing' data with analytical formula is troublesome. We therefore adopt the quadratic EOS \eqref{Quadratic_EOS} as a physically grounded and computationally convenient reference proxy. It provides a single, smooth analytical form valid from $\eta=1$, correctly reflects expected low-compression behavior via its second-order expansion, and reasonably approximates the high-density data trend seen in Fig.~\ref{fig:ExperimentalEOS_vs_formulas}.

The specific parameters used to generate the reference (quadratic EOS) profiles in Fig.~\ref{fig:Comparison} correspond to $\alpha$-phase iron (bcc), taken from Table~V in \citet{PhysRevB.108.014102}: $\rho_0=7.927$~g~cm$^{-3}$, $K=164.8$~GPa, and $K'=5.55$.

Four representative cases are presented in Fig.~\ref{fig:Comparison}. First, we consider an example with a central overdensity of 0.1\% (dilatation $\mu_c=10^{-3}$) shown in the upper-left panel (a). In this regime, all the compared models produce essentially identical, overlapping curves, demonstrating that the elastic model performs well. The total mass, $M_{(a)}=8.55 \times 10^{19}$~kg, falls in the range of objects size like Mimas or Vesta ($0.32~M_{(a)}$). The dimensionless radius, $\tilde{R}_{(a)} < 0.1$ (with $R_0\simeq 1768$~km, cf.~Table~\ref{tbl:K_rho_tbl}) is well in the elastic regime. For smaller objects, representing the majority of asteroids and smaller moons, the elastic model is therefore a very good choice, and offers an improvement over the constant density approximation.

Increasing the central overdensity to 1\% (Fig.~\ref{fig:Comparison}, panel (b)), we begin to observe the differences between the models. Two groups emerge: elastic and linear EOS tend to underestimate density (and consequently, the radius) to a similar extent, while generalized elastic EOS still matches the reference model closely.  However, discrepancy remains small. We are now in the mass range of $M_{(b)}=2.75 \times 10^{21}$~kg, comparable to objects like Ceres or Pluto ($0.21~M_{(b)}$). The elastic model remains an adequate approximation, particularly when the primary goal is to perform quick, back-of-the-envelope analytical calculations. The dimensionless radius is still less than 1 (Fig.~\ref{fig:Comparison} (b), upper-right panel, upper axis).

The third example, with an overdensity at the core of 10\%, equivalent to central dilatation $\mu_c = 10^{-1}$, brings us into the realm of larger objects.  The mass, $M_{(c)}=10^{23}$~kg, is equivalent to 0.31 of Mercury's mass, or 1.4 times the Moon's mass. Dimensionless radius $\tilde{R}_{(c)} \sim 1$ marks the approximate limit of the elastic regime. As we see in Fig.~\ref{fig:Comparison} (c) (lower-left panel, red dot-dashed curve), the elastic model begins to significantly diverge from the reference result.

Finally, we presented a case with a large central overdensity, double the uncompressed density $\rho_0$. None of the proposed analytical EOSs (\ref{ElasticEOS}, \ref{Linear_EOS}, \ref{const_Kprim}) can reproduce the reference profile obtained with the quadratic EOS \eqref{Quadratic_EOS}. This is particularly surprising for the generalized elastic EOS \eqref{const_Kprim}, which has a second-order expansion identical to that of Eq.~\eqref{Quadratic_EOS}. However, the quadratic fit \eqref{quadratic_profile} remains surprisingly accurate as evident in Fig.~\ref{fig:Comparison} (d) (blue dashed curve). This case falls within the true planetary range, with a mass of $M_{(d)}=5.17 \times 10^{24}$~kg, or 0.87 $M_\oplus$. The radius is approximately $2.5 R_0$, whereas we showed in Sect.~\ref{sec:astro} that the maximum possible radius for the elastic model is slightly below $1.5 R_0$.

We emphasize that this comparison is based on a specific study \citep{Smith2018} of the iron EOS, which is well approximated using \eqref{Quadratic_EOS}. For generic matter with arbitrary $K'$, especially for small, say $K'<4$, the range of validity might be different. Nevertheless, as a general guideline, values of dimensionless radius $\tilde{R}>1$ or dimensionless mass $\tilde{M}>1$ clearly indicate that the elastic model is no longer applicable. Typical values for a range of materials are gathered in Table~\ref{tbl:K_rho_tbl}.

In summary, even for a well-studied material like iron, the cold barotropic EOS remains remarkably complex topic and subject to significant uncertainties.  The modeling of objects such as Mercury's core (composed of up to 95\% Fe) \citep{UREY1951209,FRICKER1976,doi:10.1126/science.1218809,https://doi.org/10.1029/2020JE006792} and iron asteroids \citep{Anders1964,doi:10.1098/rsta.1988.0066,OCKERTBELL2010674} continues to be an area of active research. The theoretical challenges involved in modeling even these relatively well-understood objects are considerable. It is therefore easy to envision the difficulties that future researchers will encounter when confronted with a mysterious interstellar object,  like 'Oumuamua \citep{Drahus2018} or Borisov \citep{Jewitt_2019}, of unusual composition similar to, for instance, the suspected interstellar meteoroid Hypatia \citep{KRAMERS2022115043}, which unexpectedly enters the Solar System.  Rapidly evaluating a multitude of hypotheses in such a scenario necessitates a simple, analytical tool capable of encompassing a broad range of parameters. The elastic model, with its derived formulas, provides such a capability.

\section{Conclusions \label{sec:conclusions} }

We have derived a solution for the internal structure and properties of a perfectly elastic body characterized by a constant bulk modulus, $K$, and an initial density, $\rho_0$. All properties are determined by a single univariate special function, $U(x)$, which is the solution to the ODE given by Eq.~\eqref{elastic_equation}. This function exhibits well-defined behavior at $x \to 0$, eq.~\eqref{eq:szereg}, and $x \to \infty$, eq.~\eqref{eq:asymptotic_formula}. It can be approximated by simple analytical formulas, e.g.,  \eqref{eq:integral_von_neumann}. The properties of the model for specific astrophysical bodies are recovered through appropriate scaling of the radius \eqref{R0} and mass \eqref{M0}, and depend on a single dimensionless parameter, $\eta_c$, Eqns.~(\ref{eq:eta_c},\ref{eq:eta_c_inv}), which represents  the ratio of central density to initial density. The scaling formula \eqref{R0} for the radius is structurally identical to the well-known limits of asphericity \citep{Melosh_2011}, up to a multiplicative factor of the order of unity. However, the interpretation of the constant $K$, measured in pressure units (usually GPa), is different: it represents the bulk modulus for our model, whereas in the aspherical case it is the tensile strength. From dimensional analysis, the combination
$
K^{1/2} G^{-1/2} \rho_0^{-1}
$
is the only one that leads to the length units, cf.~\eqref{R0}.

Our theoretical analysis has uncovered a complex and nuanced behavior in the model. Initially, for radii $r \ll R_0$, the elastic model behaves as one would anticipate for incompressible matter, i.e.,  the body's size increases while maintaining a constant density. Subsequently, the effects of elasticity become significant, and the model remains a valid approximation up to a dimensionless radius of  $\tilde{R} <0.25$, corresponding to $\eta_c<1.01$. 
Beyond this point, the coefficient $K'$ can no longer be neglected, particularly for materials similar to iron, where $K' \sim 5$ (Table~\ref{tbl:K_rho_tbl}). In such cases, the elastic model breaks down when $\eta_c \gtrsim2$ (Fig.~\ref{fig:Comparison}d). However, theoretically, for hypothetical matter with $K' \ll 1$, the elastic model's applicability could extend to higher compressions, allowing us to explore its behavior even beyond. When the central density, $\rho_c$, exceeds $3 \rho_0$, the behavior of the system becomes particularly intricate.  At $\eta_c \simeq 3.562$ (Fig.~\ref{fig:MR_combined}b) the object reaches its maximum radius of $1.462 R_0$. With a further increase in mass, at  $\eta_c\simeq5.134$, the maximum mass of $4.481 M_0$ is attained. The mass-radius curve  then deflects, entering the unstable branch.  Ultimately, the mass-radius curve spirals towards the asymptotic point of the $M(R)$ relationship. This asymptotic state is approached as $\eta_c \to \infty$, and is characterized by the coordinates $R_\infty=\sqrt{2} R_0$, $M_\infty=3\sqrt{2} M_0$.

To evaluate the practical applicability of the elastic model, we employed a quadratic fit to the experimentally determined equation of state for iron. Remarkably, the simple EOS given by Eqn.~\eqref{Quadratic_EOS} reproduces the numerical data with comparable accuracy to the established (Eq.~\ref{Birch–Murnaghan}) and modern AP2 (Eq.~\ref{AP2_EOS}) analytical approximations for the iron EOS (Fig.~\ref{fig:ExperimentalEOS_vs_formulas}). This analysis  indicates that the scale $R_0$ \eqref{R0} provides a reasonable estimate for the range of validity of the elastic model, see Fig.~\ref{fig:Comparison}.

From an astronomy and astrophysics education perspective, a new analytically solvable model is invaluable in the era of  Large Language Models (LLMs). Traditional  exercises (e.g. Lane-Emden and Chandrasekhar models) are readily solved in numerous textbooks and online resources, therefore, they are already present in training data for these AI systems. These tasks can be solved instantly by any sufficiently advanced AI. This presents a growing challenge for university-level instruction, particularly for homework assignments, making them less effective as tools for genuine learning. The model presented in this work is new, and for a time, it will pose a genuine challenge to these systems, compelling students to engage in substantial intellectual effort to grasp fundamental concepts, such as the mass-radius $M(R)$ relationship, by considering readily visualized examples like liquid bodies or iron spheres (representing asteroids). We believe our article provides a refreshing bird's-eye view and serves as a modern introduction to the topic of simple, spherically symmetric, self-gravitating bodies in hydrostatic equilibrium.

Several simple formulas derived in this work, e.g., Eqs.~(\ref{quadratic_profile}-\ref{quadratic_profile:U}), could be valuable for future asteroid mining endeavors, offering analytical estimates of central compression and internal density profiles that go beyond the trivial $\rho=const$ approximation. More precise results can be obtained using Eqns.~(\ref{R}-\ref{density_contrast}).

For possible follow-up, we highlight the extraordinary effectiveness of simple quadratic EOS \eqref{Quadratic_EOS} in approximating high-pressure experimental data.  While the focus of this work was on the elastic (logotropic) model, the superior performance of the quadratic EOS in fitting real-world data suggests that the resulting ODE \eqref{Q_equation} warrants a similarly detailed mathematical investigation. Such an analysis could potentially lead to an even more accurate and widely applicable model for the internal structure of liquid/solid self-gravitating bodies. 

Future work could extend this model to piecewise EOS, with e.g. polytropic core and elastic mantle \citep{10.1093/mnras/stv1397}, which might be particularly useful for objects with homogeneous outer layer, like icy moons or exoplanets \citep{PhysRevD.104.084097}. 
Extending to rotating bodies \citep{PhysRevD.109.104049} following \citet{10.1093/mnras/93.5.390} would improve asteroid breakup predictions \citep{Drahus_2015} due to YORP effect \citep{YORP} beyond constant-density models.

\section*{Acknowledgments}
We are grateful to Prof. Piotr Bizoń for the derivation of the autonomous system equivalent to the logotropic equation, and Prof.~P.~Mach for the discussion on the existence of its solution. We also would like to thank Prof.~Ł.~Bratek for clarification on the bulk modulus of the quark matter. We also thank the anonymous referee for valuable insights.

\software{
  Julia \citep{bezanson2012julia},
  Wolfram Mathematica \citep{Mathematica}
}

\appendix

\section{Simultaneous Derivation of Elastic, Lane-Emden, and Chandrasekhar Models \label{ThreeModels}}

This appendix presents a simultaneous derivation of the internal structure and properties for an elastic model, alongside several other analytical EOSs commonly used in astrophysical modeling.The well-known Lane-Emden and Chandrasekhar theories are included for comparison, allowing the derivations to proceed in a familiar manner. This approach highlights the similarities and differences between our model and established calculations. We begin with the following EOSs:

\begin{subequations}
\begin{align}
P &= K \ln{\eta}, \quad &\text{(Elastic EOS)} \\
P &= K {\eta}^{1+1/n}, \quad &\text{(Polytropic EOS)}\\
P &= K \, f\left( {\eta}^{1/3}  \right), \quad &\text{(Chandrasekhar EOS)}\\
P &= K \, b\left( {\eta}^{1/3}  \right), \quad &\text{(Birch-Murnaghan EOS)}\\
P &= K \mu, \quad &\text{(Linear EOS)}\\
P &= K \mu - \frac{1}{2} \, K (1-K') \, \mu^2, \quad &\text{(Quadratic EOS)}\\
P &= \frac{K}{K'} \left( {\eta}^{K'} - 1 \right), \quad &\text{(Generalized Elastic EOS)}
\end{align}
\end{subequations}
where the compression, $\eta$, and dilatation, $\mu$, are defined in Eqns. \eqref{compression_ratio} and \eqref{dilatation}, respectively.
The core function for the Chandrasekhar EOS is:
$$
 f(x) = x (2x^2-3)\sqrt{1+x^2} + 3 \mathop{\mathrm{arsinh}}{x},
$$
and for the Birch-Murnaghan EOS, it is:
$$
b(x) = \frac{3}{2} \left( x^7-x^5 \right) \left[ 1+ \frac{3}{4} (K'-4)(x^2-1)\right].
$$

The specific enthalpy $h=\int dp/\rho$ is
\begin{subequations}
\begin{align}
    h &= \frac{K}{\rho_0} \left( 1 - \frac{1}{\eta} \right), \quad &\text{(Elastic)}\\
    h &=   \frac{K}{\rho_0}  (1+n) \, {\eta}^{1/n},          \quad &\text{(Polytropic)}  \\
    h &=\frac{8 K}{\rho_0} \sqrt{1+{\eta}^{2/3} },           \quad &\text{(Chandrasekhar)}\\
    h &= \frac{3}{16} \frac{K}{\rho_0}  \left(7 \eta ^{4/3} (14-3 K')+5 \eta ^{2/3} (3 K'-16)+9 \eta ^2 (K'-4)-3 (K'-6)\right), \quad &\text{(Birch-Murnaghan)}\\
    h &= \frac{K}{\rho_0} \ln{\eta}, \quad &\text{(Linear)}\\
    h &= \frac{K}{\rho_0} (K'-1) \left( \eta-1 - \frac{K'-2}{K'-1} \ln{\eta}\right), \quad &\text{(Quadratic)}\\
    h &= \frac{K}{\rho_0} \frac{{\eta}^{K'-1}-1}{K'-1}. \quad &\text{(Generalized Elastic)}
\end{align}
\end{subequations}


Before proceeding to scale and convert the hydrostatic equilibrium equations into dimensionless form, we discuss the general approach. For each EOS, three distinct approaches are possible: scaling the density by $\rho_c$ (
Case I), scaling by $\rho_0$ (
Case II), or shifting by $\rho_0$ and scaling by $\rho_c-\rho_0$ to the interval $\{0,1\}$ (
Case III). In Case I, a single family of special functions is obtained, subject to the standard initial conditions given by \eqref{initial_conditions}. The solutions are determined by the properties at the terminating point, defined by the intersection with $1/\eta_c$. Ideally, $\eta_c$ would not appear in the resulting differential equation. In Case II, a family of solutions is obtained with varying initial conditions. Finally, in Case III, all driving parameters could, in principle, appear in the ODE itself. The primary criterion for selecting among Cases I-III was to obtain the simplest form of the final ODE, minimizing the number of EOS parameters that appear within the ODE.

Substituting the following expressions for $\rho(r)$ into the hydrostatic equilibrium equation (Eq.~\ref{eq:hydro}) and Poisson's equation (or, equivalently, into Eq.~\ref{eq:LE-spherical}):
\begin{subequations}
\begin{align}
   \rho(r) &= \rho_c \, u(r/\lambda),       \quad &\text{(Elastic, Case I)}\\
   \rho(r) &= \rho_c \, \theta(r/\lambda)^n, \quad &\text{(Polytropic, Case III)}\\
   \rho(r) &=\rho_c \, z_c^3 \left( \varphi(r/\lambda)^2 - 1/z_c^2 \right)^{3/2}, \quad &\text{(Chandrasekhar, Case I)}\\
   \rho(r) &=\rho_0 \, B(r/\lambda), \quad &\text{(Birch-Murnaghan, Case II)}\\
   \rho(r) &=\rho_c \, L(r/\lambda), \quad &\text{(Linear, Case I)}\\
   \rho(r) &=\rho_0 \, Q(r/\lambda), \quad &\text{(Quadratic, Case II)}\\
   \rho(r) &=\rho_c \, R(r/\lambda), \quad &\text{(Generalized Elastic, Case I)}\\
\end{align}
\end{subequations}
where $u(x)$ is the elastic function, $\theta(x)$ - Lane-Emden polytropic function and $\varphi(x)$ - Chandrasekhar white dwarf function, $B(x)$ is Birch-Murnaghan function, $L(x)$ is the function for linear EOS, $Q(x)$ - quadratic EOS function and $R(x)$ denotes the function resulting from generalized elastic EOS. After some algebra (see attached Wolfram Language code) we obtain the following scaling relations:
\begin{subequations}
\begin{align}
    \lambda^2 &= \frac{K}{4 \pi G \rho_c^2}, \quad &\text{(Elastic)}\\
    \lambda^2 &= \frac{K(n+1)}{4 \pi G \rho_c \rho_0} \left(\rho_c/\rho_0 \right)^{1/n}, \quad &\text{(Polytropic)}\\
    \lambda^2 &= \frac{2 K}{\pi G \rho_0^2} \frac{1}{z_c^2}, \quad &\text{(Chandrasekhar)}\\
    \lambda^2 &= \frac{K}{4 \pi G \rho_0^2},                 \quad &\text{(Birch-Murnaghan)}\\
    \lambda^2 &= \frac{K}{4 \pi G \rho_0 \rho_c},            \quad &\text{(Linear)}\\
    \lambda^2 &= \frac{K}{4 \pi G \rho_0^2},                 \quad &\text{(Quadratic)}\\
    \lambda^2 &= \frac{K}{4 \pi G \rho_c^2} \left(\rho_c/\rho_0 \right)^{K'}. \quad &\text{(Generalized Elastic)}
\end{align}
\end{subequations}
The resulting dimensionless ordinary differential equations are listed below. The elastic equation is  
\begin{subequations}
\makeatletter\@fleqntrue\makeatother
\begin{flalign}
    &\Delta u - \frac{2 u'^2}{u} + u^3=0. \quad &\text{(Elastic)}
\end{flalign}
Lane-Emden polytropic equation \citep{Horedt2004} is
\begin{flalign}
\label{Lane-Emden}
    &\Delta \theta + \theta^n = 0. \quad &\text{(Polytropic)}
\end{flalign}
Chandrasekhar's white dwarf equation \citep[37.1 Chandrasekhar’s Theory]{Kippenhahn2012} is
\begin{flalign}
    &\Delta \varphi + \left( \varphi^2 - \frac{1}{z_c^2} \right)^{3/2} = 0. \quad &\text{(Chandrasekhar)}
\end{flalign}
Equation related to Birch-Murnaghan EOS is too complicated, so we skip it. 
\addtocounter{equation}{1}
For linear EOS \eqref{Linear_EOS} we get
\begin{flalign}
    &\Delta L - \frac{{L'}^2}{L} + L^2 = 0. \quad &\text{(Linear)}
\end{flalign}
For quadratic EOS \eqref{Quadratic_EOS} we got 
\begin{flalign}
\label{Q_equation}
    &\Delta Q + \frac{(1-\delta)Q^3+\delta {Q'}^2}{Q(Q-\delta)}=0, \qquad \delta = \frac{K'-2}{K'-1}. \quad &\text{(Quadratic)}
\end{flalign}
Finally, generalized elastic EOS leads to
\begin{flalign}
    &\Delta R + (K'-2) \frac{{R'}^2}{R}+R^{3-K'}=0. \quad &\text{(Generalized Elastic)}
\end{flalign}
\makeatletter\@fleqnfalse\makeatother
\end{subequations}
As intended, setting $K'=0$ in the equation for the generalized elastic EOS recovers the elastic equation. Furthermore, in the limit $K'\to1$ both the generalized elastic and quadratic EOSs reduce to the linear EOS, demonstrating that the linear EOS is a special case of both.

The initial conditions for the respective ODEs are:
\begin{subequations}
\begin{alignat}{3}
u(0)       &= 1,       \;&u'(0)&=0,       \quad \text{(Elastic)}\\
\theta(0)  &= 1,       \;&\theta'(0)&=0,  \quad \text{(Polytropic)}\\
\varphi(0) &= 1,       \;&\varphi'(0)&=0, \quad \text{(Chandrasekhar)} \\
B(0)       &= \eta_c,  \;&B'(0)&=0,       \quad \text{(Birch-Murnaghan)}\\
L(0)       &= 1,       \;&L'(0)&=0,       \quad \text{(Linear)}\\
Q(0)       &= \eta_c,  \;&Q'(0)&=0,       \quad \text{(Quadratic)}\\
R(0)       &= 1,       \;&R'(0)&=0.       \quad \text{(Generalized Elastic)}
\end{alignat}
\end{subequations}

The integration is terminated at a dimensionless radius  $x_t$, defined by the following conditions:
\begin{subequations}
\begin{align}
u(x_t)        & = 1/\eta_c, \quad &\text{(Elastic)}\\
\theta(x_t)   & = 0,        \quad &\text{(Polytropic)}\\
\varphi(x_t)  & = 1/z_c,    \quad &\text{(Chandrasekhar)}\\
B(x_t)        & = 1,        \quad &\text{(Birch-Murnaghan)}\\
L(x_t)        & = 1/\eta_c, \quad &\text{(Linear)}\\
Q(x_t)        & = 1,        \quad &\text{(Quadratic)}\\
R(x_t)        & = 1/\eta_c, \quad &\text{(Generalized Elastic)}
\end{align}
\end{subequations}
respectively.

\bibliography{ElasticPlanetoids}{}
\bibliographystyle{aasjournal}

\end{document}